%% file: RJwrapper.tex
\documentclass[a4paper]{report}
\usepackage[utf8]{inputenc}
\usepackage[T1]{fontenc}
\usepackage{RJournal}
\usepackage{amsmath,amssymb,array}
\usepackage{booktabs}

\usepackage{threeparttable}
\usepackage{booktabs}
\usepackage{longtable}
\usepackage{adjustbox}
\usepackage{lscape}
\usepackage{enumitem}

\begin{document}

\sectionhead{Contributed research article}
\volume{XX}
\volnumber{YY}
\year{20ZZ}
\month{AAAA}

\begin{article}
  \input{did2s}
\end{article}

\end{document}

%% file: did2s.tex
\title{\pkg{did2s}: Two-Stage Difference-in-Differences}
\author{by Kyle Butts and John Gardner}

\maketitle

\abstract{%
Recent work has highlighted the difficulties of estimating
difference-in-differences models when the treatment is adopted at
different times for different units. This article introduces the R
package \CRANpkg{did2s} which implements the estimator introduced in
\citet{Gardner_2021}. The article provides an approachable review of the
underlying econometric theory and introduces the syntax for the function
\texttt{did2s}. Further, the package introduces functions, event\_study
and plot\_event\_study, which uses a common syntax to implement all of
the modern event-study estimators.
}

\hypertarget{introduction}{%
\subsection{Introduction}\label{introduction}}

A rapidly growing econometric literature has identified difficulties in
traditional difference-in-differences estimation when treatment turns on
at different times for different groups and when the effects of
treatment vary across groups and over time
\citep{Callaway_SantAnna_2020, Sun_Abraham_2020, Goodman-Bacon_2018, Borusyak_Jaravel_Spiess_2021, deChaisemartin_DHaultfoeuille_2019}.
\citet{Gardner_2021} proposes an estimator of the two-way fixed-effects
model that is quick and intuitive. The estimator relies on the standard
two-way fixed-effect model (see the following section) and forms an
intuitive estimate: the average difference in outcomes between treated
and untreated units after removing fixed unit- and time-invariant
shocks.

This article first discusses the modern difference-in-differences theory
in an approachable way and second discusses the software package,
\CRANpkg{did2s}, which implements the two-stage estimation approach
proposed by \citet{Gardner_2021} to estimate robustly the two-way
fixed-effects (TWFE) model. There are two notable technical features of
this package. First, \texttt{did2s} utilizes the incredibly fast
package, \CRANpkg{fixest} \citep{Berge_2018}, which can estimate
regressions with a high number of fixed-effects very quickly. Since
there are a few alternative TWFE event-study estimators implemented in
\texttt{R}, each with their own syntax and data formatting requirements,
the package also has a set of functions that allow quick estimation and
plotting of every alternative event study estimator using a standardized
syntax. This allows for easy comparison between the results of different
methods.

\hypertarget{difference-in-differences-theory}{%
\subsection{Difference-in-Differences
Theory}\label{difference-in-differences-theory}}

Researchers commonly use the difference-in-differences (DiD) methodology
to estimate the effects of treatment in the case where treatment is
non-randomly assigned. Instead of random assignment giving rise to
identification, the DiD method relies on the so-called ``parallel
trends'' assumption, which asserts that outcomes would evolve in
parallel between the treated and untreated groups \emph{in a world where
the treated were untreated}. This is formalized with the \emph{two-way
fixed-effects} (TWFE) model. In a static setting where treatment effects
are \emph{constant} across treatment groups and over time, researchers
estimate the \textbf{static TWFE model}: \begin{equation}\label{eq:twfe}
  y_{igt} = \mu_g + \eta_t + \tau D_{gt} + \varepsilon_{igt},
\end{equation} where \(y_{igt}\) is the outcome variable of interest,
\(i\) denotes the individual, \(t\) denotes time, and \(g\) denotes
group membership where ``group'' is defined as all units that start
treatment at time \(g\).\footnote{in the literature, never treated units
  often are given a value of \(g = \infty\).} \(\mu_g\) is a vector of
time-invariant group characteristics, \(\eta_t\) is a vector of shocks
in a given time period that is experienced by all individuals equally,
and \(D_{gt}\) is an indicator for whether initial-treatment group \(g\)
is receiving treatment in period \(t\),
i.e.~\(D_{gt} \equiv \mathbb{1}(g \leq t)\). The coefficient of interest
is \(\tau\), which is the \textbf{(\emph{constant}) average effect of
the treatment on the treated} (ATT). If it is indeed true that the
treatment effect is constant across groups and over time, then the
estimate formed by estimating the static TWFE model will be consistent
for \(\tau\).

However, treatment effects are not constant in most settings. The
magnitude of a unit's treatment effect can differ based on group status
\(g\) (e.g.~if groups that benefit more from a policy implement it
earlier) and treatment duration (e.g.~if treatment effects grow as the
policy has been in place for longer periods). Therefore to enrich our
model, we allow heterogeneity in treatment effects across \(g\) and
\(t\) by introducing the \textbf{group-time average treatment effect},
\(\tau_{gt}\). Correspondingly, we modify the TWFE model as follows: \[
  y_{igt} = \mu_g + \eta_t + \tau_{gt} D_{gt} + \varepsilon_{igt}.
\] The key difference is that treatment effects are allowed to differ
based on group status \(g\) and time period \(t\). Estimating any
individual \(\tau_{gt}\) may not be desirable since there would be too
few observations. Instead, researchers aggregate group-time average
treatment effects into the \textbf{overall average treatment effect},
\(\tau\), which averages across \(\tau_{gt}\): \[
  \tau \equiv \sum_{g, t} \frac{1}{N_{gt}} \tau_{gt},
\] where \(N_{gt}\) denotes the number of (post-treatment) group-time
pairs, \(\{ g, t \}\), observed in our sample. The natural question is,
``does the static TWFE model produce a consistent estimate for the
overall average treatment effect?'' Except for a few specific scenarios,
the answer is no
\citep{Sun_Abraham_2020, Goodman-Bacon_2018, Borusyak_Jaravel_Spiess_2021, deChaisemartin_DHaultfoeuille_2019}.

One way of thinking about this disappointing result is through the
Frisch--Waugh--Lovell (FWL) theorem \citep{Frisch_Waugh_1933}. This
theorem says that estimating the Static TWFE model is equivalent to
estimating \[
y_{igt} - \hat{\mu}_g - \hat{\eta}_t = \tau \tilde{D}_{gt} + \tilde{\varepsilon}_{gt},
\] where \(\tilde{D}_{gt}\) denotes the residuals from regressing
\(D_{gt}\) on \(\mu_g\) and \(\eta_t\), \(\hat{\mu}_g\) and
\(\hat{\eta}_t\) are estimates for the group and time fixed-effects,
respectively. The left-hand side of this equation, under a parallel
trends restriction on the error term \(\varepsilon_{it}\), is our
estimate for \(\tau_{gt}\). Therefore, the FWL theorem tells us
estimating the static TWFE model is equivalent to estimating\footnote{This
  is a minor abuse of notation since
  \(y_{igt} - \hat{\mu}_g - \hat{\eta}_t\) is an estimate for
  \(\tau_{igt}\) which can be different from \(\tau_{gt}\) if there is
  within group-time heterogeneity.}

\[
\hat{\tau}_{gt} = \tau \tilde{D}_{gt} + \tilde{\varepsilon}_{gt}
\]

The resulting estimate for \(\tau\) can be written as: \[
\hat{\tau} \equiv \sum_{g,t} w_{gt} \hat{\tau}_{gt},
\] where \(w_{gt}\) is the weight put on the corresponding
\(\hat{\tau}_{gt}\). Results of \citet{Gardner_2021},
\citet{Borusyak_Jaravel_Spiess_2021}, and
\citet{deChaisemartin_DHaultfoeuille_2019} all characterize the weights
\(w_{gt}\) from this regression. There are only two cases where the
\(\hat{\tau}\) is a consistent estimate for the overall average
treatment effect. First, when treatment occurs at the \emph{same time}
for all treated units, then \(w_{gt}\) is equal to \(1/N_{gt}\) for all
\(\{g, t\}\) and therefore \(\hat{\tau}\) is a consistent estimate for
the overall average treatment effect. The other scenario when
\(\hat{\tau}\) estimates the overall average treatment effect is when
\(\tau_{gt}\) is constant across group and time,
i.e.~\(\tau_{gt} = \tau\). Since the weights, \(w_{gt}\), always sum to
one, we have that
\(\hat{\tau} = \sum w_{gt} \hat{\tau}_{gt} \to \sum w_{gt} \tau = \tau\).

The above cases are not the norm in research. If there is heterogeneity
in group-time treatment effects \emph{and} differences in when units get
treated, then \(\hat{\tau}\) is not a consistent estimate for the
average treatment effect \(\tau\). Instead, \(\hat{\tau}\) will be a
weighted average of group-time treatment effects with some weights,
\(w_{gt}\), being potentially negative. This yields a treatment effect
estimate that does not provide a good summary of the ``average''
treatment effect. It is even possible for the sign of \(\hat{\tau}\) to
differ from that of the overall average treatment effect. This would
occur, for example, if negative weights are placed primarily on the
largest (in magnitude) group-time treatment effects.

To summarize the modern literature, the fundamental problem faced in
estimating the TWFE model is the potential negative weighting. The
proposed methodology in \citet{Gardner_2021} is based on the fact that
if \(\hat{\tau}_{gt}\) is regressed on \(D_{gt}\), instead of
\(\tilde{D}_{gt}\), the resulting weights would be exactly equal to
\(1/N_{gt}\) and the coefficient of \(D_{gt}\) would estimate the
overall average treatment effect.

\hypertarget{event-study-estimates}{%
\subsubsection{Event-study Estimates}\label{event-study-estimates}}

Researchers have attempted to model treatment effect heterogeneity by
allowing treatment effects to change over time. To do this, they
introduce a (dynamic) event-study TWFE model: \begin{equation}
  y_{igt} = \mu_g + \eta_t + \sum_{k = -L}^{-2} \tau^k D_{gt}^k + \sum_{k = 1}^{K} \tau^k D_{gt}^k + \varepsilon_{igt},
\end{equation} where \(D_{gt}^k\) are lags/leads of treatment (\(k\)
periods from initial treatment date). The coefficients of interests are
the \(\tau^k\), which represent the average effect of being treated for
\(k\) periods. For negative values of \(k\), \(\tau^k\) are known as
``pretrends,'' and represent the average deviation in outcomes for
treated units \(k\) periods away from treatment, relative to their value
in the reference period. These pre-trend estimates are commonly used as
a test of the parallel counterfactual trends assumption.

Our goal is to estimate the \textbf{average treatment effect of being
exposed for \(k\) periods}, an average of \(\tau_{gt}\) for only the set
of \(\{g,t\}\) where \(k\) periods have elapsed since \(g\),
i.e.~\(t - g = k\): \[
  \tau^k = \sum_{g,t \ : \ t - g = k} \frac{1}{N_{gt}^k} \tau_{gt}
\] where the sum is over \(\{g,t\}\) with \(t - g = k\) and \(N_{gt}^k\)
is the count of \(\{g,t\}\) pairs that satisfy that condition. The
results of \citet{Sun_Abraham_2020} show that even though we allow for
our average treatment effects to vary over time \(\tau^k\), the negative
weighting problems would arise if units are treated at different times
\emph{and} there is group-heterogeneity in treatment effects. Similar to
the static TWFE model, the estimates of \(\tau^k\) from running the
event-study model form non-intuitively weighted averages of
\(\tau_{gt}\) with \(w_{gt}^k \neq N_{gt}^k\). Even worse, the
group-time treatment effects for \(t-g \neq k\) will be included in the
estimate of \(\hat{\tau}^k\). Hence, the need for a robust
difference-in-differences estimator remains even in the event-study
model.

\hypertarget{two-stage-difference-in-differences-estimator}{%
\subsubsection{Two-stage Difference-in-Differences
Estimator}\label{two-stage-difference-in-differences-estimator}}

\citet{Gardner_2021} proposes an estimator to resolve the problem with
the two-way fixed-effects approaches. Rather than attempting to estimate
the group and time effects simultaneously with the ATT (causing
\(D_{it}\) to be residualized), Gardner's approach proceeds from the
observation that, under parallel trends, the group and time effects are
identified from the subsample of untreated/not-yet-treated observations
(\(D_{gt} = 0\)). This suggests a simple two-stage
difference-in-differences estimator:

\begin{enumerate}
\def\labelenumi{\arabic{enumi}.}
\item
  Estimate the model \[
   y_{igt} = \mu_g + \eta_t + \varepsilon_{igt}
  \] using the subsample of untreated/not-yet-treated observations
  (i.e., all observations for which \(D_{gt}=0\)), retaining the
  estimated group and time effects to form the adjusted outcomes
  \(\tilde{y}_{igt} \equiv y_{igt} - \hat{\mu}_g - \hat{\eta}_t\).
\item
  Regress adjusted outcomes \(\tilde{y}_{igt}\) on treatment status
  \(D_{gt}\) or \(D_{gt}^k\) in the full sample to estimate treatment
  effects \(\tau\) or \(\tau^k\).
\end{enumerate}

To see why this procedure works, note that parallel trends implies that
outcomes can be expressed as \begin{align*}
  y_{igt} &= \mu_g + \eta_t + \tau_{gt} D_{gt} + \varepsilon_{igt} \\
  &= \mu_g + \eta_t + \bar{\tau} D_{gt} + (\tau_{gt} - \bar{\tau}) D_{gt} + \varepsilon_{igt},
\end{align*} where \(\tau_{gt} = E(Y^1_{igt} - Y^0_{igt} \ | \ g, t)\)
is the average treatment effect for group \(g\) in period
\(t\)\footnote{i.e., the average difference between treated and
  untreated potential outcomes \(y^1_{igt}\) and \(y^0_{igt}\),
  conditional on the observed treatment-adoption times.} and
\(\bar{\tau} = E(\tau_{gt} | D_{gt}=1)\) is the overall average
treatment effect\footnote{i.e., the population-weighted average of the
  group-time specific ATTs, \(\tau_{gt}\).}. Note from parallel trends,
\(E(\varepsilon_{igt} | D_{gt}, g, t) = 0\). Rearranging, this gives \[
  y_{igt} - \mu_g - \eta_t = \bar{\tau} D_{gt} + (\tau_{gt} - \bar{\tau}) D_{gt} + \varepsilon_{igt}.
\] Suppose you knew the time and group fixed-effects and were able to
directly observe the left-hand side (later we will estimate the
left-hand side). Regressing the adjusted \(y\) variable, on \(D_{gt}\)
will produce a consistent estimator for \(\bar{\tau}\). To see this,
note that \(E[(\tau_{gt} - \bar{\tau}) D_{gt} \ | \ D_{gt}] = 0\).
Hence, the treatment dummy is uncorrelated with the omitted variable and
the average treatment effect is identified in the second-stage. Since we
are not able to directly observe \(\mu_g\) and \(\eta_t\), we estimate
them using the untreated/not-yet-treated observations in the
first-stage. However, standard errors need adjustment to account for the
added uncertainty from the first-stage estimation.

This approach can be extended to dynamic models by replacing the second
stage of the procedure with a regression of residualized outcomes onto
the leads and lags of treatment status, \(D_{gt}^k\),
\(k \in \{-L, \dots, K\}\). Under parallel trends, the second-stage
coefficients on the lags identify the overall average effect of being
treated for \(k\) periods (where the average is taken over all units
treated for at least that many periods). The second-stage coefficients
on the leads identify the average deviation from predicted
counterfactual trends among units that are \(k\) periods away from
treatment, which under parallel trends should be zero for any
pre-treatment value of \(k\). Hence, the coefficients on the leads
represent a test of the validity of the parallel trends assumption.

\hypertarget{inference}{%
\subsubsection{Inference}\label{inference}}

The standard variance-covariance matrix from the second-stage regression
will be incorrect since it fails to account for the fact that the
dependent variable is generated from the first-stage regression.
However, this estimator takes the form of a joint generalized method of
moments (GMM) estimator whose asymptotic variance is well understood
\citep{newey_mcfadden_1994}.

Specifically, the estimator takes the form of a two-stage GMM estimator
with the following two moment conditions: \begin{align}
  m(\theta) = (Y-X_{10}'\gamma)X_{10} \\
  g(\gamma, \theta) = (Y - X_1'\gamma - X_2'\theta) X_2,
\end{align} where \(X_1\) is the matrix of group and time fixed-effects,
\(X_{10}\) corresponds to the matrix \(X_1\), but with rows
corresponding to observations for which \(D_{gt} = 1\) replaced with
zeros (as only observations with \(D_{gt} = 0\) are used in the first
stage) and \(X_2\) is the matrix of treatment variable(s). The first
equation corresponds with the first stage and the second equation
corresponds with the second stage. From Theorem 6.1 of
\citet{newey_mcfadden_1994}, the asymptotic variance of the two-stage
estimator is \begin{equation}
  V = G_\theta^{-1} E\left[ (g + G_\gamma \psi)(g + G_\gamma \psi)' \right] G_\theta^{-1'},
\end{equation} where from our moment conditions, we have: \[
  G_\theta = - E\left(X_2X_2' \right),
\] \[
  G_\gamma = - E\left(X_2X_1'\right),
\] \[
  \psi = E(X_{10}X_{10}')^{-1} \varepsilon_{10} X_{10}.
\]

This can be estimated using \begin{equation}
  \left(X_2'X_2\right)^{-1} \left(\sum_{g=1}^G W_g' W_g\right) \left(X_2'X_2\right)^{-1},
\end{equation} where \[
  W_g = X_{2g}'\hat{\varepsilon}_{2g} - \hat{\varepsilon}_{10g}' X_{1g}\left(X_{1g}'X_{1g}\right)^{-1} \left(X_{1g}'X_{2g}\right)
\] and matrices indexed by \(g\) correspond to the \(g\)th cluster.

\hypertarget{the-package}{%
\subsection{\texorpdfstring{The \CRANpkg{did2s}
Package}{The  Package}}\label{the-package}}

The \CRANpkg{did2s} package introduces two sets of functions. The first
is the \texttt{did2s} command which implements the two-stage
difference-in-differences estimator as described above. The second is
the \texttt{event\_study} and \texttt{plot\_event\_study} commands that
allow individuals to implement alternative `robust' estimators using a
singular common syntax.

\hypertarget{the-did2s-command}{%
\subsubsection{\texorpdfstring{The \texttt{did2s}
Command}{The did2s Command}}\label{the-did2s-command}}

The command \texttt{did2s} implements the two-stage
difference-in-differences estimator following \citet{Gardner_2021}. The
general syntax is

\begin{Schunk}
\begin{Sinput}
did2s(data, yname, first_stage, second_stage, 
      treatment, cluster_var, weights = NULL, 
      bootstrap = FALSE, n_bootstraps = 250, 
      verbose = TRUE)
\end{Sinput}
\end{Schunk}

and full details on the arguments is available in the help page,
available by running \texttt{?did2s}. There are a few arguments that are
worth discussing in more detail.

The \texttt{first\_stage} and \texttt{second\_stage} arguments require
formula arguments. These formulas are passed to the
\texttt{fixest::feols} function from \CRANpkg{fixest} and can therefore
utilize two non-standard formula options that are worth mentioning
\citep{Berge_2018}. First, fixed-effects can be inserted after the
covariates,
e.g.~\texttt{\textasciitilde{}\ x1\ \textbar{}\ fe\_1\ +\ fe\_2}, which
will make estimation much faster than using \texttt{factor(fe\_1)}.
Second, the function \texttt{fixest::i} can be used for treatment
indicators instead of \texttt{factor}. The advantage of this is that you
can easily specify the reference values, e.g.~for event-study indicators
where researchers typically want to drop time \(t = -1\),
\texttt{\textasciitilde{}\ i(rel\_year,\ ref\ =\ c(-1))} would be the
correct second-stage formula. Additionally, \CRANpkg{fixest} has a
number of post-estimation exporting commands to make tables with
\texttt{fixest::etable} and event-study plots with
\texttt{fixest::iplot}/\texttt{fixest::coefplot}. The \texttt{fixest::i}
function is better integrated with these functions as we will see below.

The option \texttt{treatment} is the variable name of a \(0/1\) variable
that denotes when treatment is active for a given unit, \(D_{gt}\) in
the above notation. Observations with \(D_{gt} = 0\) will be used to
estimate the first stage, which removes the problem of treatment effects
contaminating estimation of the unit and time fixed-effects. However, as
an important note, if you suspect anticipation effects before treatment
begins, the \texttt{treatment} variable should be shifted forward by
\(x\) periods for observations to prevent the aforementioned
contamination. For example, if you suspect that units could experience
treatment effects 1 period ahead of treatment (a so-called anticipatory
effect), then the \texttt{treatment} should begin one period ahead.
These anticipation effects can be estimated, after adjusting the
\texttt{treatment} variable, by using a reference year of say,
\(t = -2\) and looking at the estimate for relative year \(-1\).

\hypertarget{example-usage-of-did2s}{%
\paragraph{\texorpdfstring{Example usage of
\texttt{did2s}}{Example usage of did2s}}\label{example-usage-of-did2s}}

For basic usage, I will use the simulated dataset, \texttt{df\_het},
that comes with the \CRANpkg{did2s} package with the command

\begin{Schunk}
\begin{Sinput}
data(df_het, package = "did2s")
\end{Sinput}
\end{Schunk}

The data-generating process is displayed in Figure \ref{fig:ex-data}.
The lines represent the mean outcome for each treatment group and the
never-treated group. In the absence of treatment, each group is
simulated to be on parallel trends. There is heterogeneity in treatment
effects both within a treatment group over time and across treatment
groups.

\begin{Schunk}
\begin{figure}
\includegraphics[width=1\linewidth]{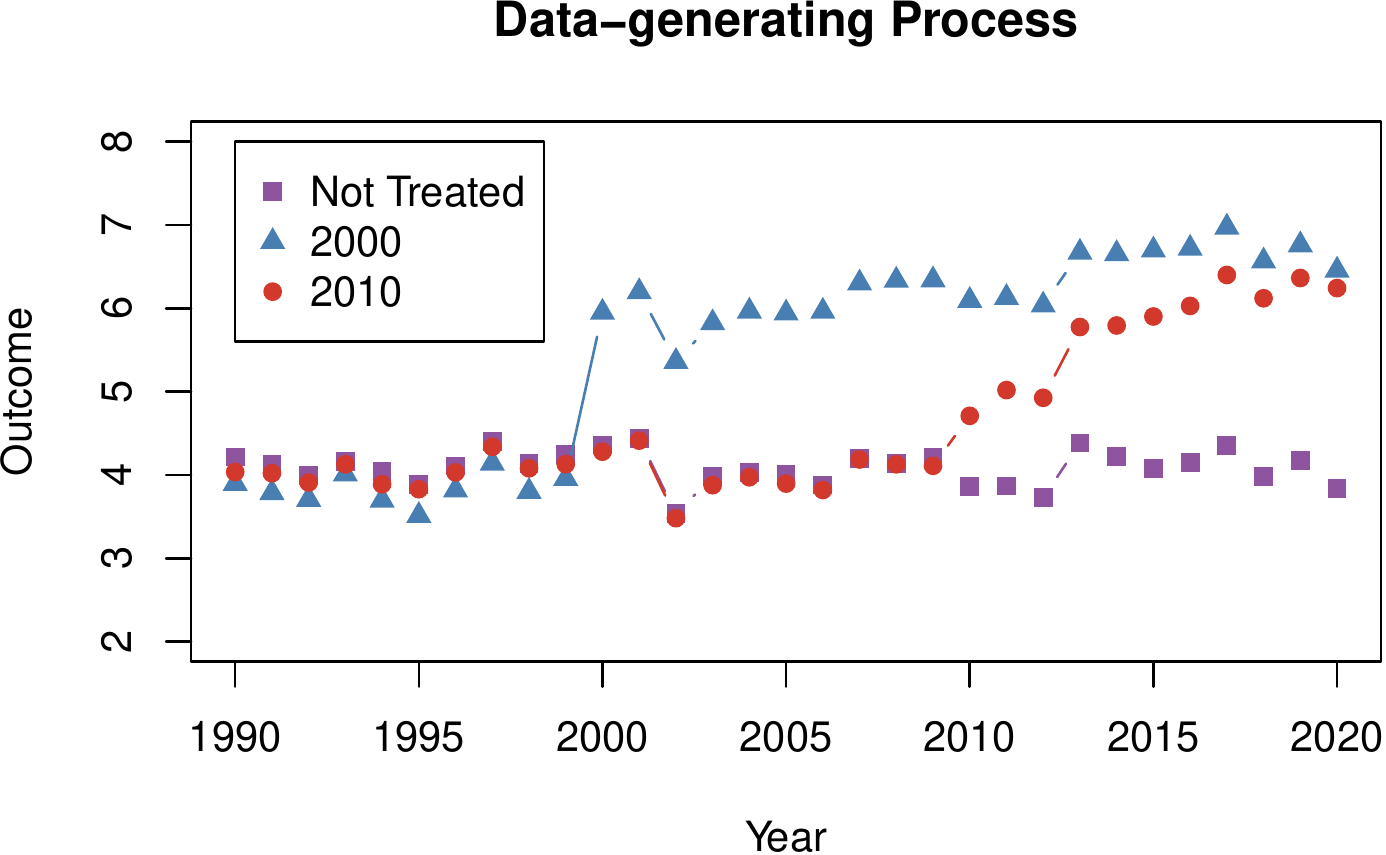} \caption[Example data with heterogeneous and dynamic treatment effects]{Example data with heterogeneous and dynamic treatment effects. Each line represents the average outcome in a given year for each group. In the absence of treatment, all three groups would exhibit parallel trends.}\label{fig:ex-data}
\end{figure}
\end{Schunk}

First, we will calculate a static difference-in-differences estimate
using the \texttt{did2s} function.

\begin{Schunk}
\begin{Sinput}
static = did2s(
    data = df_het, 
    yname = "dep_var", 
    treatment = "treat",
    first_stage = ~ 0 | unit + year, 
    second_stage = ~ i(treat, ref = FALSE),
    cluster_var = "unit", 
    verbose = FALSE
)

summary(static)
\end{Sinput}
\begin{Soutput}
#> OLS estimation, Dep. Var.: dep_var
#> Observations: 155,000 
#> Standard-errors: Custom 
#>             Estimate Std. Error t value  Pr(>|t|)    
#> treat::TRUE  2.25957   0.011705 193.037 < 2.2e-16 ***
#> ---
#> Signif. codes:  0 '***' 0.001 '**' 0.01 '*' 0.05 '.' 0.1 ' ' 1
#> RMSE: 1.04088   Adj. R2: 0.513594
\end{Soutput}
\end{Schunk}

Since the returning object is a \texttt{fixest} object, all the
accompanying output commands from \CRANpkg{fixest} are available to use.
For example, we can create regression tables:

\begin{Sinput}
fixest::etable(static, fitstat = c("n"), tex = TRUE,
               title = "Estimate of Static TWFE Model", 
               notes = "Standard errors clustered at unit level. 
Estimated using Two-Stage Difference-in-Differences.
proposed by Gardner (2021).")
\end{Sinput}
\begin{table}[htbp]
   \caption{Estimate of Static TWFE Model}
   \centering
   \begin{tabular}{lc}
      \tabularnewline \midrule \midrule
      Dependent Variable: & dep\_var\\   
      Model:              & (1)\\  
      \midrule
      \emph{Variables}\\
      treat $=$ TRUE      & 2.260$^{***}$\\   
                          & (0.0117)\\   
      \midrule
      \emph{Fit statistics}\\
      Observations        & 155,000\\  
      \midrule \midrule
      \multicolumn{2}{l}{\emph{Custom standard-errors in parentheses}}\\
      \multicolumn{2}{l}{\emph{Signif. Codes: ***: 0.01, **: 0.05, *: 0.1}}\\
   \end{tabular}
   
   \par \raggedright 
   Standard errors clustered at unit level. 
   Estimated using Two-Stage Difference-in-Differences.
   proposed by Gardner (2021).
\end{table}

However, since there are dynamic treatment effects in this example, it
is much better to estimate the dynamic effects themselves using an
event-study specification. We will then plot the results using
\texttt{fixest::iplot}, which plots coefficients corresponding to an
\texttt{i()} variable. Note that \texttt{rel\_year} is coded as
\texttt{Inf} for never-treated units, so this has to be noted in the
reference part of the formula.

\begin{Schunk}
\begin{Sinput}
es = did2s(
    data = df_het, 
    yname = "dep_var", 
    treatment = "treat",
    first_stage = ~ 0 | unit + year, 
    second_stage = ~ i(rel_year, ref = c(-1, Inf)),
    cluster_var = "unit", 
    verbose = FALSE
)

fixest::iplot(
    es, 
    main = "Event study: Staggered treatment", 
    xlab = "Relative time to treatment", 
    col = "steelblue", ref.line = -0.5
)

# Add the (mean) true effects
true_effects = tapply((df_het$te + df_het$te_dynamic), df_het$rel_year, mean)
true_effects = head(true_effects, -1)
points(-20:20, true_effects, pch = 20, col = "grey60")

# Legend
legend(x=-20, y=3, col = c("steelblue", "grey60"), 
       pch = c(20, 20), 
       legend = c("Two-stage estimate", "True effect"))
\end{Sinput}
\begin{figure}
\includegraphics[width=1\linewidth]{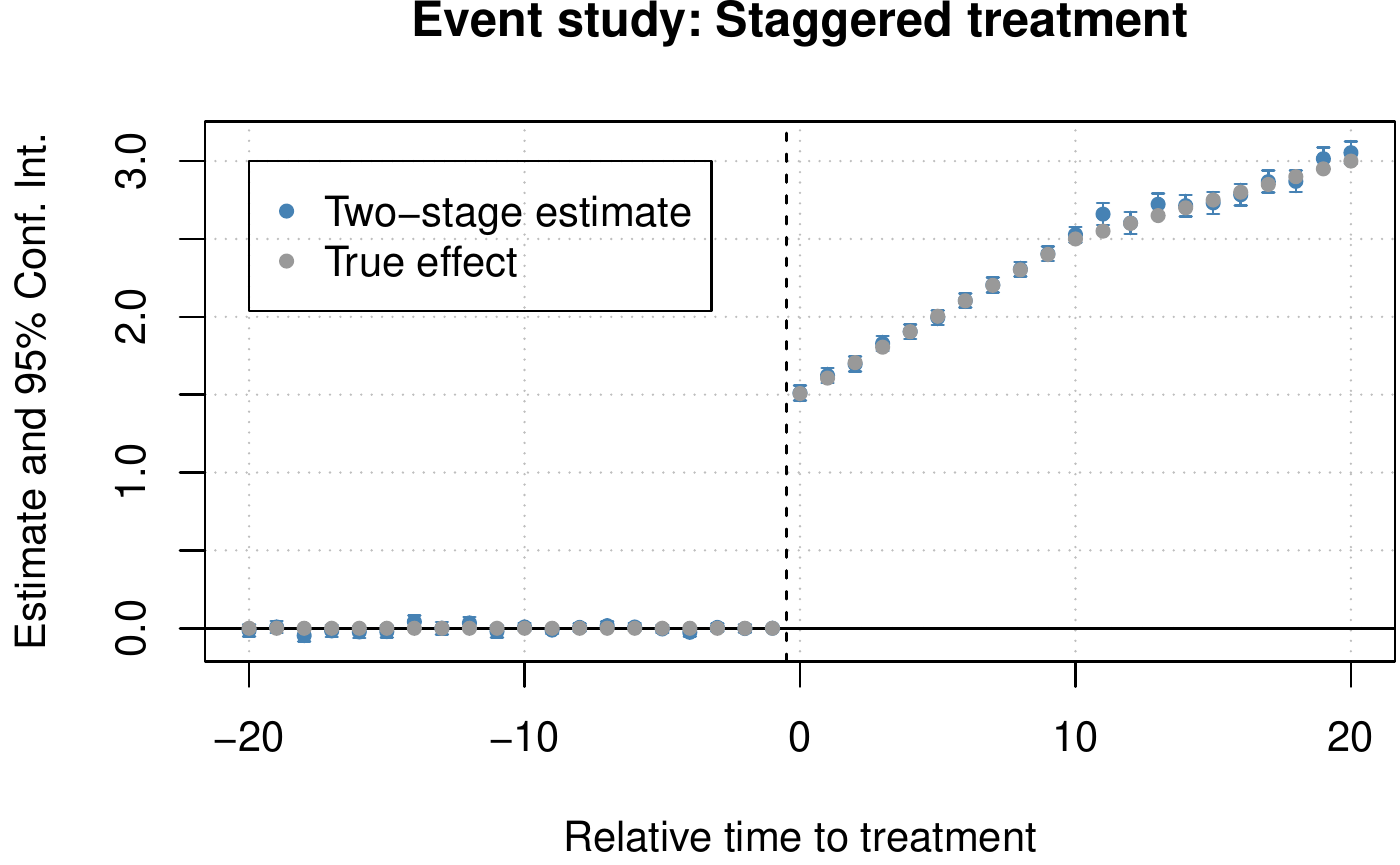} \caption[Event-study Estimate of TWFE Model]{Event-study Estimate of TWFE Model. Standard Errors clustered at unit level. Estimated using the Two-Stage Difference-in-Differences proposed by Gardner (2021).}\label{fig:dynamic}
\end{figure}
\end{Schunk}

The event study estimates are found in Figure \ref{fig:dynamic} and
match closely to the true average treatment effects. For comparison to
traditional OLS estimation of the event-study specification, Figure
\ref{fig:dynamic-w-twfe} plots point estimates from both methods. As
pointed out by \citet{Sun_Abraham_2020}, treatment effect heterogeneity
between groups biases the estimated pre-trends. In the figure below, the
OLS estimates appear to show violations of pre-trends even though the
data was simulated under parallel pre-trends.

\begin{Schunk}
\begin{Sinput}
twfe = feols(dep_var ~ i(rel_year, ref=c(-1, Inf)) | unit + year, data = df_het) 

fixest::iplot(list(es, twfe), sep = 0.2, ref.line = -0.5,
      col = c("steelblue", "#82b446"), pt.pch = c(20, 18), 
      xlab = "Relative time to treatment", 
      main = "Event study: Staggered treatment (comparison)")

# True Effects
points(-20:20, true_effects, pch = 20, col = "grey60")

# Legend
legend(x=-20, y=3, col = c("steelblue", "#82b446", "grey60"), pch = c(20, 18, 20), 
       legend = c("Two-stage estimate", "TWFE", "True Effect"))
\end{Sinput}
\begin{figure}
\includegraphics[width=1\linewidth]{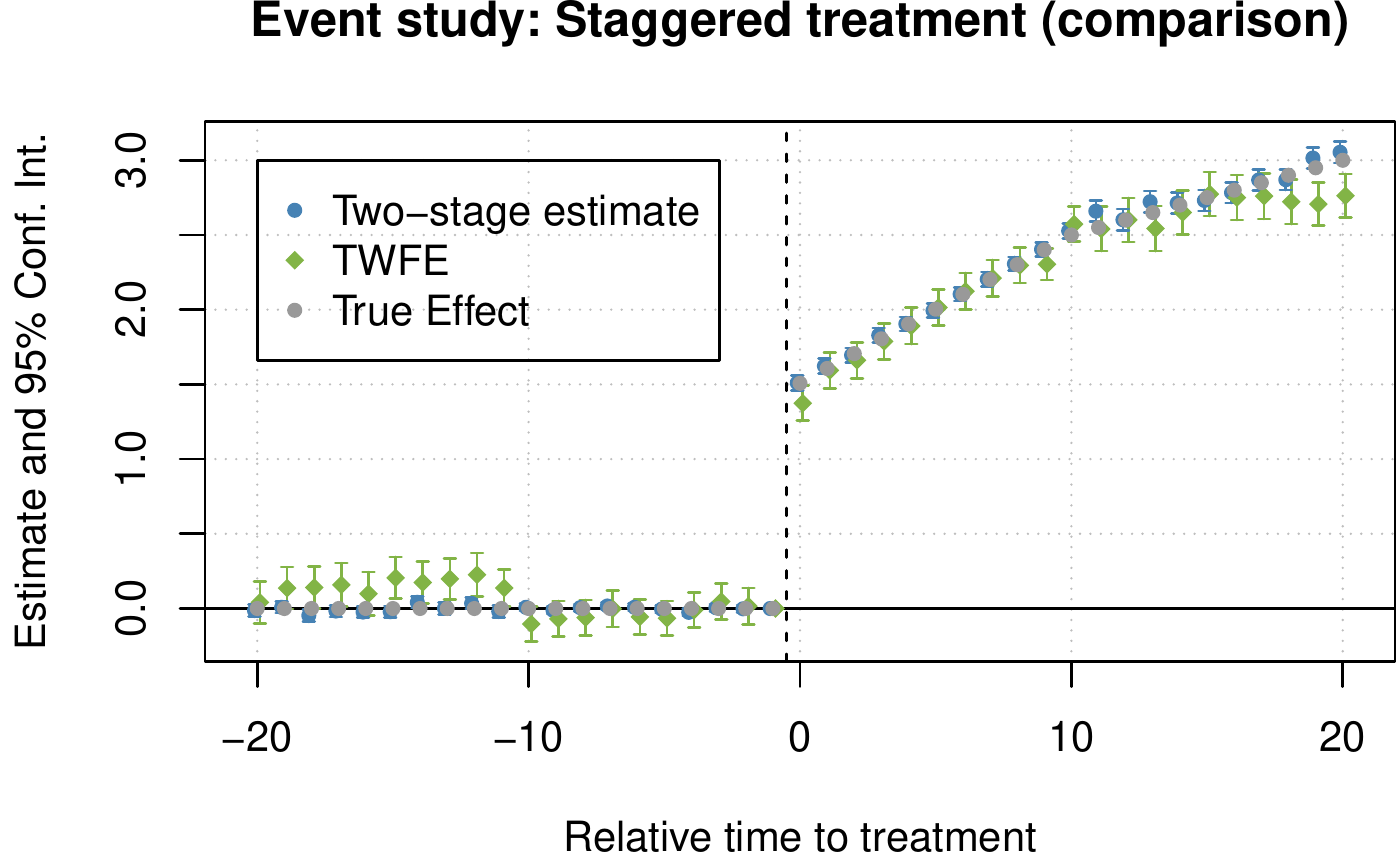} \caption[Event-study Estimate of TWFE Model]{Event-study Estimate of TWFE Model. Standard Errors clustered at unit level. Estimated using the Two-Stage Difference-in-Differences proposed by Gardner (2021) and a Traditional TWFE model.}\label{fig:dynamic-w-twfe}
\end{figure}
\end{Schunk}

\hypertarget{the-event_study-and-plot_event_study-command}{%
\subsubsection{\texorpdfstring{The \texttt{event\_study} and
\texttt{plot\_event\_study}
command}{The event\_study and plot\_event\_study command}}\label{the-event_study-and-plot_event_study-command}}

The command \texttt{event\_study} presents a common syntax that
estimates the event-study TWFE model for treatment-effect heterogeneity
robust estimators recommended by the literature and returns all the
estimates in a data.frame for easy plotting by the command
\texttt{plot\_event\_study}. The general syntax is

\begin{Schunk}
\begin{Sinput}
event_study(
    data, yname, idname, tname, gname, 
    estimator,
    xformla = NULL, horizon = NULL, weights = NULL
)
\end{Sinput}
\end{Schunk}

The option \texttt{data} specifies the data set that contains the
variables for the analysis. The four other required options are all
names of variables: \texttt{yname} corresponds with the outcome variable
of interest; \texttt{idname} is the variable corresponding to the
(unique) unit identifier, \(i\); \texttt{tname} is the variable
corresponding to the time period, \(t\); and \texttt{gname} is a
variable indicating the period when treatment first starts (group
status).

\begin{landscape}
\begin{table}
\caption{Event Study Estimators \label{tab:estimators-latex}}
\begin{threeparttable}

\begin{tabular}{p{0.2\textwidth}llp{0.2\textwidth}p{0.4\textwidth}c}
\toprule
\multicolumn{1}{l}{Estimator} & \multicolumn{1}{l}{R package /} & \multicolumn{1}{l}{Type} & \multicolumn{1}{l}{Comparison group} & \multicolumn{1}{l}{Main Assumptions} & \multicolumn{1}{l}{Uniform inference} \\
 & \multicolumn{1}{l}{\texttt{estimator} argument} & & & & \\

\midrule
Gardner (2021) 
& \texttt{did2s} 
& Imputes $Y(0)$ 
& Not-yet- and/or Never-treated 
& 
\vspace{-5mm}
\begin{itemize}[leftmargin=*]
  \item Parallel Trends for all units
  \item Limited anticipation${^*}$
  \item Correct specification of $Y(0)$
\end{itemize}
& \\

Borusyak, Jaravel, and Spiess (2021) 
& \texttt{didimputation} 
& Imputes $Y(0)$ 
& Not-yet- and/or Never-treated 
& 
\vspace{-5mm}
\begin{itemize}[leftmargin=*]
  \item Parallel Trends for all units
  \item Limited anticipation${^*}$
  \item Correct specification of $Y(0)$
\end{itemize}
& \\

Callaway and Sant'Anna (2021) 
& \texttt{did} 
& 2$\times$2 Aggregation 
& Either Not-yet- or Never-treated 
& 
\vspace{-5mm}
\begin{itemize}[leftmargin=*]
  \item Parallel Trends for Not-yet-treated {\it or} Never-treated
  \item Limited anticipation${^*}$
\end{itemize}
& Yes \\

Sun and Abraham (2020) 
& \texttt{fixest}/\texttt{sunab}
& 2$\times$2 Aggregation 
& Not-yet- and/or Never-treated 
& 
\vspace{-5mm}
\begin{itemize}[leftmargin=*]
  \item Parallel Trends for all units
  \item Limited anticipation${^*}$
\end{itemize}
& \\

Roth and Sant'Anna (2021) 
& \texttt{staggered} 
& 2$\times$2 Aggregation 
& Not-yet-treated 
& 
\vspace{-5mm}
\begin{itemize}[leftmargin=*]
  \item Treatment timing is random
  \item Limited anticipation${^*}$
\end{itemize}
& \\
\bottomrule
\end{tabular}

\begin{tablenotes}
\item This table summarizes the differences between various proposed event-study estimators in the econometric literature.
\item[$^{*}$]  Anticipation can be accounted for by adjusting 'initial treatment day' back $x$ periods, where $x$ is the number of periods before treatment that anticipation can occur.
\end{tablenotes}

\end{threeparttable}
\end{table}
\end{landscape}

There are five main estimators available and the choice is specified for
the \texttt{estimator} argument and are described in Table
\ref{tab:estimators-latex}.\footnote{Except for Sun and Abraham, the
  \texttt{estimator} option is the package name. For Sun and Abraham,
  the \texttt{estimator} option is \texttt{sunab}. A value of ``all''
  will estimate all 5 estimators.} The following paragraphs will aim to
highlight the differences and commonalities between estimators. These
estimators fall into two broad categories of estimators. First,
\CRANpkg{did2s} and \CRANpkg{didimputation} \citep{didimputation} are
\texttt{imputation-based} estimators as described above. Both rely on
``residualizing'' the outcome variable
\(\tilde{Y} = Y_{it} - \hat{\mu}_g - \hat{\eta}_t\) and then averaging
those \(\tilde{Y}\) to estimate the event-study average treatment effect
\(\tau^k\). These two estimators return identical point estimates, but
differ in their asymptotic regime and hence their standard errors.

The second type of estimator, which we label \texttt{2x2\ aggregation},
takes a different approach for estimating event-study average treatment
effects. The packages \CRANpkg{did} \citep{did}, \CRANpkg{fixest} and
\CRANpkg{staggered} \citep{staggered} first estimate \(\tau_{gt}\) for
all group-time pairs. To estimate a particular \(\tau_{gt}\), they use a
two-period (periods \(t\) and \(g-1\)) and two-group (group \(g\) and a
``control group'') difference-in-differences estimator, known as a
\texttt{2x2} difference-in-differences. The particular ``control group''
they use will differ based on estimator and is discussed in the next
paragraph. Then, the estimator manually aggregate \(\tau_{gt}\) across
all groups that were treated for (at least) \(k\) periods to estimate
the event-study average treatment effect \(\tau^k\).

These estimators do not all rely on the same underlying assumptions, so
the rest of the table summarizes the primary differences between
estimators. The comparison group column describes which units are
utilized as comparison groups in the estimator and hence will determine
which units need to satisfy a parallel trends assumption. For example,
in some circumstances, treated units will look very different from
never-treated units. In this case, parallel trends may only hold between
ever-treated units and hence only these units should be used in
estimation. In other cases, for example if treatment is assigned
randomly, then it's reasonable to assume that both not-yet- and
never-treated units would all satisfy parallel trends.

For estimators labeled ``Not-yet- and/or never-treated'', the default is
to use both not-yet- and never-treated units in the estimator. However,
if all never-treated units are dropped from the data set before using
the estimator, then these estimators will use only not-yet-treated
groups as the comparison group. \CRANpkg{did} provides an option to use
either the not-yet- treated \emph{or} the never- treated group as a
comparison group depending on which group a researcher thinks will make
a better comparison group. \CRANpkg{staggered} will automatically drop
units that are never treated from the sample and hence only use
not-yet-treated groups as a comparison group.

The next column, \texttt{Main\ Assumptions}, summarize concisely the
main theoretical assumptions underlying each estimator. First, the
assumptions about parallel trends match the previous discussion on the
correct comparison group. The only estimator that doesn't rely on a
parallel trends assumption is \CRANpkg{staggered} which relies on the
assumption that \emph{when} a unit receives treatment is random.

The next assumption, that is common across all estimators, is that there
should be ``limited anticipation'' of treatment. In general,
anticipatory effects are when units respond to treatment before it is
\emph{actually} implemented. For example, this can be common if the news
of a policy triggers behavior responses before the treatment is put in
place. ``Limited anticipation'' is when these anticipatory effects can
only exist in a ``few'' pre-periods.\footnote{There should be more
  periods before treatment in the sample than whatever number a ``few''
  is.} In any of these cases, ``treatment'' should be manually moved
back by the maximum number of periods where anticipation can occur. For
example, if treatment starts in 2012 and anticipatory effects are
reasonably only possible 2 years before, this units' ``group'' should be
labeled as 2010 in the data.

The \texttt{imputation-based} estimators require an additional
assumption that the parametric model of
\(Y(0) = \mu_i + \eta_t + \varepsilon_{it}\) is correctly specified.
This is because in the first stage, you have to accurately impute
\(Y(0)\) when residualizing \(Y\) which relies on the correct
specification of \(Y(0)\). The \texttt{2x2\ aggregation} models do not
estimate a parametric form of \(Y(0)\) and hence only relies on a
parallel trends assumption. The last column highlights that
\CRANpkg{did} allows for uniform inference of estimates. This addresses
the problem that multiple hypotheses tests are being done by researchers
(e.g.~checking individually if all post periods are significant) by
creating standard errors that adjust for multiple testing.

\hypertarget{example-usage-of-event_study}{%
\paragraph{\texorpdfstring{Example usage of
\texttt{event\_study}}{Example usage of event\_study}}\label{example-usage-of-event_study}}

The result of \texttt{event\_study} is a tibble in a \texttt{tidy}
format \citep{broom} that contains point estimates and standard errors
for each relative time indicator for each estimator. The results of
\texttt{event\_study} are stored as a dataframe with event-study term,
the estimate, standard error, and a column containing which estimator is
used for that estimate. This output dataframe will in turn be passed to
\texttt{plot\_event\_study} for easy comparison.
\texttt{plot\_event\_study} will return a \texttt{ggplot} object
\citep{ggplot2}. We return to the \texttt{df\_het} dataset to see
example usage of these functions.

\begin{Schunk}
\begin{Sinput}
data(df_het, package = "did2s")
out = event_study(
  data = df_het, yname = "dep_var", idname = "unit",
  tname = "year", gname = "g", estimator = "all"
)
\end{Sinput}
\begin{Soutput}
#> Estimating TWFE Model
\end{Soutput}
\begin{Soutput}
#> Estimating using Gardner (2021)
\end{Soutput}
\begin{Soutput}
#> Estimating using Callaway and Sant'Anna (2020)
\end{Soutput}
\begin{Soutput}
#> Estimating using Sun and Abraham (2020)
\end{Soutput}
\begin{Soutput}
#> Estimating using Borusyak, Jaravel, Spiess (2021)
\end{Soutput}
\begin{Soutput}
#> Estimating using Roth and Sant'Anna (2021)
\end{Soutput}
\begin{Sinput}

head(out)
\end{Sinput}
\begin{Soutput}
#>    estimator  term   estimate  std.error
#>       <char> <num>      <num>      <num>
#> 1:      TWFE   -20 0.04097725 0.07167704
#> 2:      TWFE   -19 0.13665695 0.07147683
#> 3:      TWFE   -18 0.14015820 0.07245520
#> 4:      TWFE   -17 0.15793252 0.07431871
#> 5:      TWFE   -16 0.09910002 0.07379570
#> 6:      TWFE   -15 0.20561127 0.07116478
\end{Soutput}
\end{Schunk}

\begin{Schunk}
\begin{Sinput}
plot_event_study(out, horizon = c(-5,10))
\end{Sinput}
\begin{figure}
\includegraphics[width=1\linewidth]{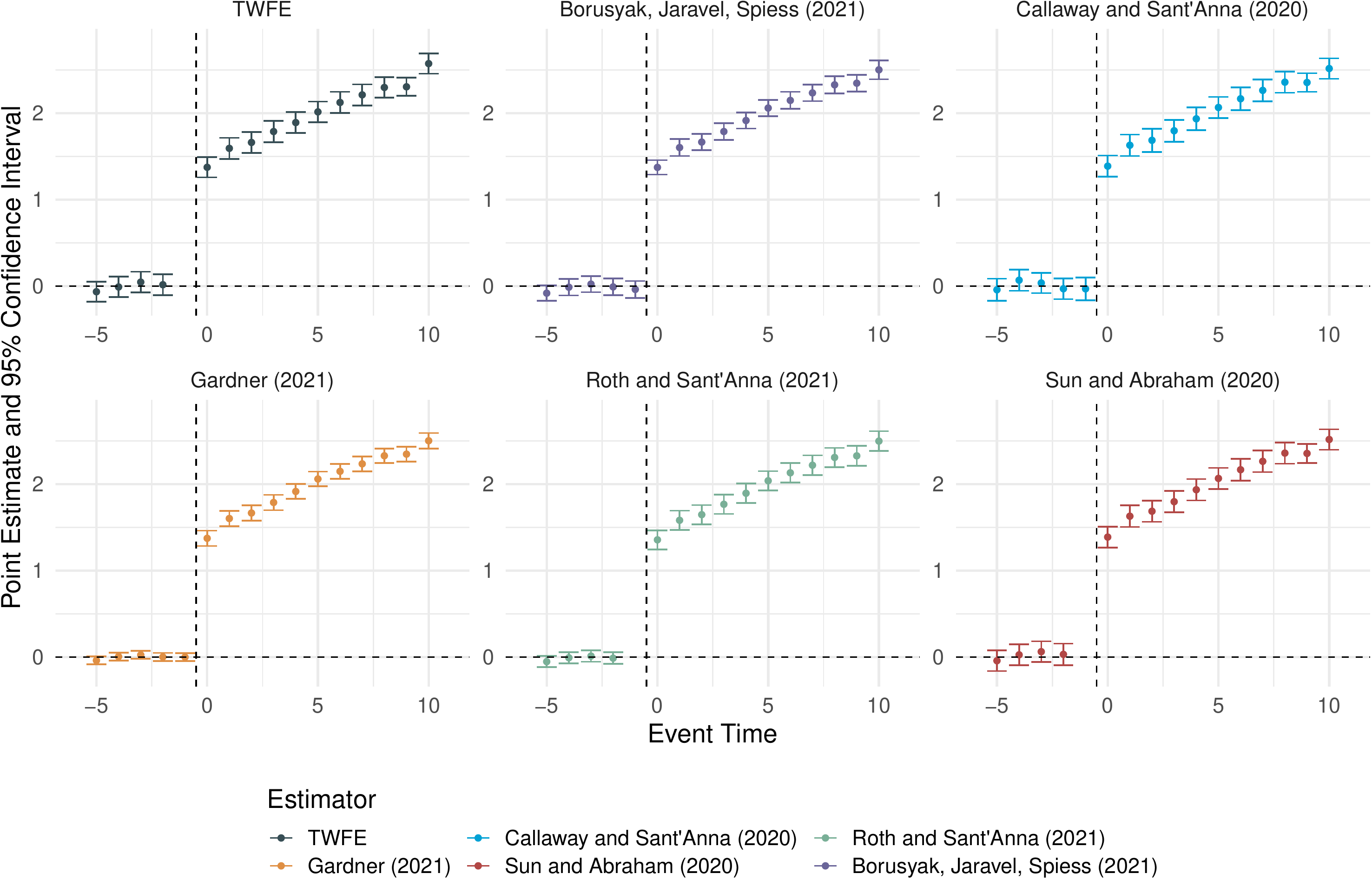} \caption[Event-study Estimators]{Event-study Estimators. This figure plots the results from the event\_study command on example data.}\label{fig:es-alternatives}
\end{figure}
\end{Schunk}

\hypertarget{conclusion}{%
\subsection{Conclusion}\label{conclusion}}

This article introduced the package \CRANpkg{did2s} which provides a
fast, memory-efficient, and treatment-effect heterogeneity robust way to
estimate two-way fixed-effect models. The package also includes the
\texttt{event\_study} and \texttt{plot\_event\_study} functions to allow
for a single syntax for the various estimators introduced in the
literature. A companion package in Stata is also available with similar
syntax for the \texttt{did2s} function.

While this package includes an event\_study function that aims to help
individuals implement any of the proposed modern ``solutions'' to the
difference-in-differences estimation, further research on this topic is
needed to help practitioners be able to more precisely determine which
estimators work best in their settings. Potentially, there could be
data-driven methods to try to identify the plausibility of the different
assumptions. Additionally, there is still more work to be done to
formalize under what conditions covariates can flexibly be used in
estimation. There is some initial work from
\citet{Caetano_Callaway_Payne_Rodrigues_2022}, but there does not yet
exist statistical software to perform their proposed estimator.

\bibliography{did2s.bib}

\address{%
Kyle Butts\\
University of Colorado Boulder\\%
\\
\url{https://www.kylebutts.com/}\\%
\textit{ORCiD: \href{https://orcid.org/0000-0002-9048-8059}{0000-0002-9048-8059}}\\%
\href{mailto:buttskyle96@gmail.com}{\nolinkurl{buttskyle96@gmail.com}}%
}

\address{%
John Gardner\\
University of Mississippi\\%
\\
\url{https://jrgcmu.github.io/}\\%
\href{mailto:jrgardne@olemiss.edu}{\nolinkurl{jrgardne@olemiss.edu}}%
}